# Upconversion of terahertz phonons in spin-ladder compounds via nonlinear coupling

S. J. Sreerag[1], N. Malathi[1], Shahla Yasmin Mathengattil[1], Rabindranath Bag[2], Surjeet Singh[2] and Rajeev N. Kini[1*]

[1]Indian Institute of Science Education and Research Thiruvananthapuram (IISER TVM), Maruthamala P.O. Vithura, Kerala 695551, India

[2]Indian Institute of Science Education and Research Pune, Dr Homi Bhabha Road, Pashan, Pune - 411008, India

*Email: rajeevkini@iisertvm.ac.in

We demonstrate dynamic control of the lattice by THz light by exploiting the coupling of phonon modes. The low-energy sliding phonon modes in the spin-ladder system, $Sr_{14}Cu_{24}O_{41,}$ are excited using THz radiation with high electric fields. Due to the nonlinearities induced by the THz electric fields, the low-energy phonon mode couples to the higher energy, silent optical phonon modes at ≈ 1.17 THz. This indirect excitation of the silent phonon mode is reflected as an enhancement of the THz transmission near 1.17 THz. Our results demonstrate that it is possible to indirectly control inaccessible phonon modes with THz frequency electromagnetic radiation, which provides opportunities to alter such material systems' electronic and magnetic properties dynamically using THz radiation.

# Introduction

Optical spectroscopy using intense THz pulses has emerged as a promising tool to study collective excitations in condensed matter systems. THz measurements have been successfully employed to study superconducting compounds that have gaps in the THz region.[1–6] Usually, a pump-probe scheme is employed to probe the effects of the intense THz pulse: a probe pulse is used to measure the pump-induced changes to the transmitted or reflected probe pulse as a function of pump-probe time delay. Phonons, collective modes like the soft phonon modes in charge density wave (CDW) systems, collective charge, and order parameter oscillations in superconductors (SCs) give rise to coherent oscillations of the probe field in the pump-probe experiments.[1–4,6–9] Ordinary THz transmission experiments at high fields are closely interconnected with the pump-probe measurements, and the observed effects in both cases are due to a strong resonance at a particular frequency caused by optical nonlinearities.[10] In complex condensed matter systems, drastic changes in the electronic and magnetic properties can be obtained from subtle changes in its crystal lattice because electronic correlations make the collective properties of such solids a highly nonlinear (NL) function of the perturbation.[11] Modifying the crystal lattice structure is conventionally accomplished by changing the chemical composition, temperature, pressure, strain, or magnetic field. For dynamically changing the crystal lattice structure and thus the electronic properties, ultrafast optical pulses in the visible to near-infrared (NIR) frequencies are used. The coupling of light with crystal lattice has been experimentally demonstrated, primarily through three means: (i) direct coupling of infrared (IR) active modes, (ii) excitation of the electronic system using pulses in the NIR or visible region, and indirect coupling of the electronic system to the lattice via electron-phonon interaction and (iii) excitation of an IR active phonon mode using mid-infrared (MIR) radiation and coupling via lattice anharmonicity. Recent advances in MIR laser techniques have opened a new field called 'nonlinear phononics', which uses resonant excitation of phonon modes, allowing us to control indirectly the otherwise inaccessible phonon modes.[12][13] Apart from that, intense broadband THz radiation has also been shown to alter the lattice structure by driving a low-frequency phonon mode into a strongly NL regime.[14] In such a case, it was shown that the low-frequency phonon could couple to a phonon mode of higher frequency.

Here, we demonstrate an upconversion process where the lower energy phonon modes, driven by THz radiation, couple to higher energy, silent phonon modes in the spin-ladder compound, $Sr_{14}Cu_{24}O_{41}$ (SCO) at high THz electric field (kV/cm) intensities. SCO consists of both spin chains and spin ladders, which are stacked alternatively along the *b*-axis. A layer of Sr atoms separates the two sublayers. The chain and ladder sub-lattices are incommensurate along the *c*-axis ($c_{ladder} \approx \sqrt{2}\, c_{chain}$), and lattice parameters are identical ($a_{ladder} = a_{chain}$) along the rung direction (*a*-axis). From valence counting, one can see that SCO is intrinsically hole-doped, leaving the Cu atoms in an average valence state of +2.25. These holes mainly reside in the chain layer with a small fraction in the ladder layer, and below the charge ordering temperature, the holes in the ladders form pairs along the rung direction (*a*-axis).[15–17] The hole pairs in the ladders crystallize and give rise to a CDW below the temperature, $T_{CDW}$. [15,16].

Due to the large number (316) of atoms in the unit cell, SCO's infrared and optical properties are extremely complicated, with several optically active phonon modes.[18] Low-frequency translational modes of the Sr, Cu, O ions, stretching, and breathing modes of the Cu-O bonds could be observed in the range of 3 – 20 THz. [18] The sliding motion of the chain and ladder planes along the incommensurate (*c*-axis) direction gives rise to the pseudo-acoustic phonon modes, which were observed in the 250 – 400 GHz range. [19,20] Unlike the conventional approach where high-frequency modes are resonantly driven by radiation in the mid-infrared to infrared region, here we show that higher energy modes indirectly couple to other lower energy phonon modes when driven into an NL regime. [14] This observation further opens a window into the opportunities to control quantum materials through dynamical modification of lattice structure via coupled phonon modes.

## The Experiment

The experiments were performed on high-quality single crystals grown using the traveling-solvent floating-zone technique. The crystal growth and characterization details can be found in Ref. [21]. The current study's central part was conducted on an SCO crystal doped with 1% Ni (SCNO sample). Measurements were also conducted on an undoped SCO crystal, which was annealed for 36 hours at 850 °C in an $O_2$ atmosphere. The crystals were cleaved along the *a-c* plane into 500 µm thin slices and cut into approximately 4.5 mm × 3 mm pieces. Terahertz time-domain spectroscopy (THz-TDS) measurements were carried out in two different systems. For low field (linear) measurements, THz signals were generated by photoexcitation of a low-temperature-grown GaAsBi epilayer using ultrafast (≈ 80 fs), NIR (λ ≈ 800 nm, 0.1 nJ) pulses from a Ti: Sapphire oscillator (repetition rate = 80 MHz).[22] The THz signals were then detected using a photoconductive antenna, and we estimated the peak electric fields to be in the range of a few tens of mV/cm. Since the low field THz pulses were generated using a high repetition rate laser, measurements using this had a better signal-to-noise ratio compared to the high field measurements. For the high field (NL) measurements, the THz signals were generated using a $LiNbO_3$ crystal, employing the tilted-pulse-front technique[23], and the transmitted THz pulses were detected by electro-optic sampling in 0.2 mm thick, (110) ZnTe crystal. The $LiNbO_3$ crystal was pumped with 88 µJ pulses from a Ti-Sapphire amplifier (repetition rate = 1 KHz), and we estimate the peak electric field to be about 6 kV/cm. For the measurements reported here, the THz electric field ($E$) was oriented parallel to the crystals' c-axis ($E||c$). THz transmittance was obtained by taking the ratio of the amplitudes of the fast Fourier-transformed sample-transmitted THz signals and the reference THz signals. THz experiments were done by mounting the crystals on a sample holder with an aperture of ≈ 3 - 3.5 mm diameter, then attaching it to the cold finger of a closed-cycle cryostat with THz-transparent windows. The THz experimental set-ups were purged with nitrogen gas to avoid absorption of THz radiation by atmospheric moisture/water vapour.

## Results and Discussion

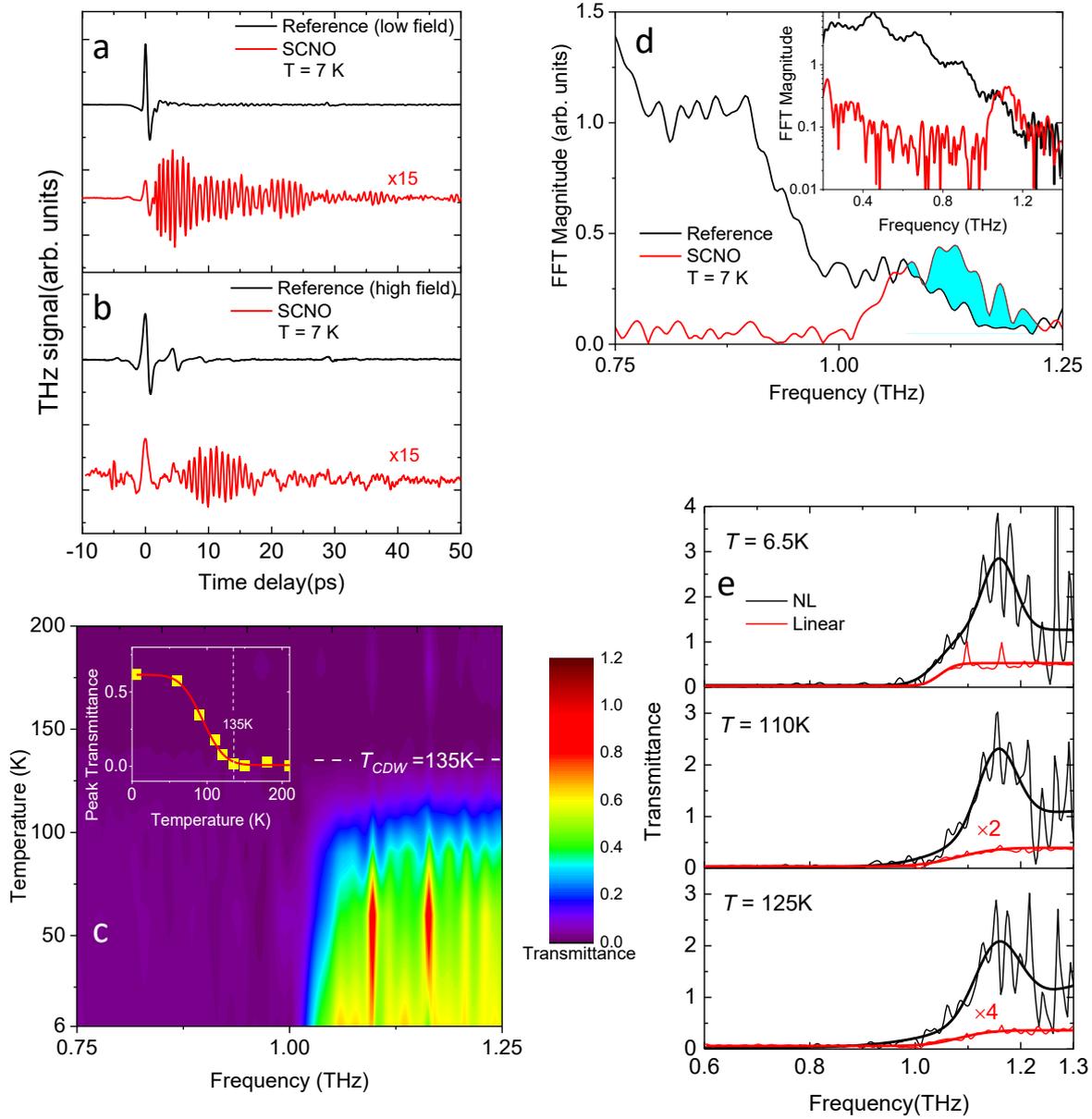

**Figure 1.** (a) The THz signal generated using GaAsBi emitter for the linear measurements and the signal transmitted through the SCO sample doped with 1% Ni (SCNO) at 7 K. (b) The THz signal generated using the tilted-pulse-front technique in LiNbO$_3$ for the nonlinear (NL) measurements, and the signal transmitted through the SCNO sample at 7 K. (c) The temperature-dependent transmittance spectra in the linear regime for the SCNO sample. The inset shows the transmittance near 1.16 THz as a function of temperature. The onset of the THz transmission is used to determine the $T_{CDW}$. (d) The FFT magnitude of the reference signal for the NL measurements and the signal transmitted through the SCNO crystal at 7 K. The inset shows the same data on a larger scale (log-linear plot). (e) A comparison of the linear (red lines) and nonlinear transmittance (thin black lines) of the SCNO crystal at 6.5 K, 110 K, and 125 K. The thick black and red lines represent a fit to the mode with a function described in the main text.

Figure 1(a) shows the transmitted THz time-domain signal through the SCNO sample at $T \approx 7$ K (linear regime) along with the reference THz pulse. The reference pulse was obtained by recording the THz pulse transmitted through an identical hole in the sample holder. For the measurements with the THz electric field ($E$) parallel to the c-axis ($E||c$), in agreement with previous reports on SCO, we observed a strong oscillatory component in the transmitted signal with a period of $\approx 1$ ps, which was thought to be arising due to the CDW phason mode.[20,24] Similar to previous reports on SCO, the transmission of THz electromagnetic radiation with frequencies below $\approx 1$ THz is highly attenuated through the SCNO [see Fig. 1(c)] at low temperatures. It has been reported earlier that the sharp cut-off near 1 THz is observed only below the CDW transition temperature, $T_{CDW}$. From the temperature dependence of the transmittance, we obtain $T_{CDW} \approx 135$ K for the SCNO sample, as shown in the inset of Figure 1(c). Compared to the undoped SCO samples in which $T_{CDW} \approx 162$ K was reported earlier, $T_{CDW}$ is lower, possibly due to Ni doping.[24]

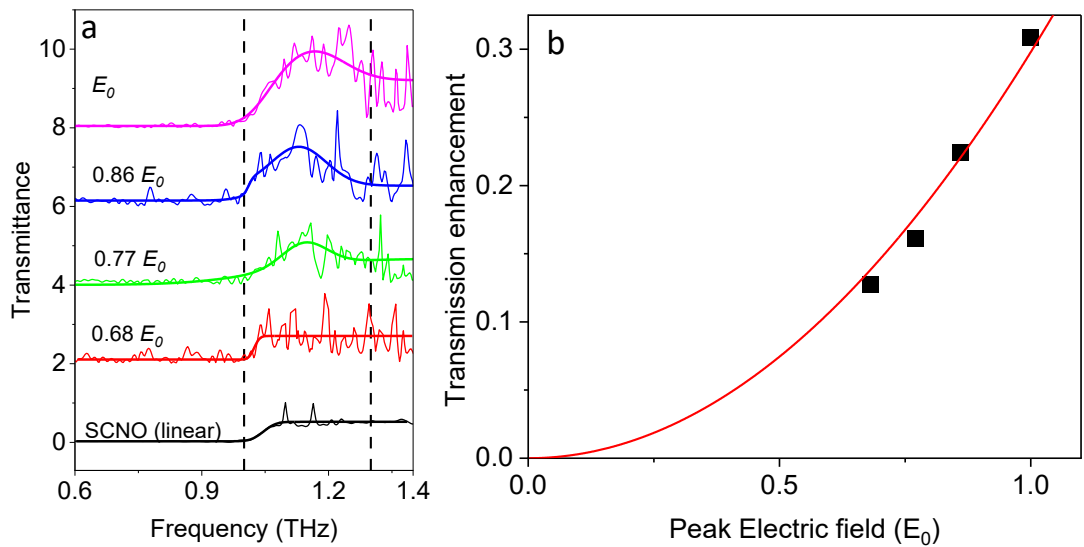

Figure 2. (a) The transmittance of SCNO sample at 7K at different THz peak electric fields. $E_0$ corresponds to 6 kV/cm. (b) The transmission enhancement at $\approx 1.17$ THz is plotted as a function of the THz pulse peak electric field. The red line is a quadratic fit to the data.

This report focuses on THz studies on SCNO, using intense THz radiation generated by the tilted-phase front technique in LiNbO$_3$. At higher excitation electric field intensities ($E_0 \approx 6$ kV/cm), the transmitted THz signal shows oscillations, as seen in figure 1(b), and a high pass cut-off near 1 THz, as shown in figure 1(d), similar to what was observed in the linear excitation case. However, in the case of intense THz excitation, an enhancement in the magnitude of the transmitted THz signal, near 1.17 ± 0.07 THz, was observed. This enhancement can be understood from the shaded region in Figure 1(d), where the FFT magnitude of the reference signal and the signal transmitted through the SCNO sample are compared.

SCO has sliding phonon modes in the 0.25 – 0.4 THz region, and hence, these modes are resonantly excited by the intense THz radiation. [19,20,24,25] The spectral content of the driving THz field near 1.17 THz is very low. But as can be seen from Fig 1(d), the magnitude of the transmitted field is larger than the driving THz magnitude near 1.17 THz. To better visualize this, the THz transmittance through SCNO, at different temperatures, under low and high excitation electric field intensities is shown in Figure 1(e). Due to the cut-off near 1 THz, the low-temperature transmittance, in the frequency range $\approx$ 0.5 – 1.3 THz, in the linear regime could be approximated to a step (Gauss error) function, as shown in Figure 1(e).

The NL transmittance could be approximated to the sum of a step function and a Gaussian function. The enhancement in the THz transmittance near 1.17 THz could be observed only at high THz peak fields and at low temperatures ($T$ < 135 K), which points to a phenomenon occurring below $T_{CDW}$. Further, this suggests that the phenomenon is nonlinear in nature, which occurs only above a certain threshold field or depends nonlinearly on the excitation THz field. To confirm this, we performed measurements at different THz electric field intensities. To change the intensity of the THz electric field, we used a neutral density filter to alter the intensity of the NIR laser pulse, which generated THz radiation in the LiNbO$_3$ crystal through the tilted-pulse-front technique. Supplementary Figure S1 shows the spectra of the driving THz field with different peak electric fields and spectral content of the driving THz field between 0.25 – 0.45 THz and 1.05 – 1.25 THz range, estimated by integrating the area of the spectrum in this frequency range. It can be seen that spectral content in the low-frequency (0.25 – 0.45 THz) region increases almost linearly with the increase in peak electric field, whereas the spectral content in the high-frequency (1.05 – 1.25 THz) region is low and does not vary significantly with the increase in peak electric field. The transmittance of the SCNO sample at 7K at different THz peak electric fields is shown in Figure 2 (a), and it can be seen that the peak transmittance near 1.17 THz increases as the excitation THz electric field increases. We estimated the transmission enhancement near the 1.17 THz peak by calculating the increase in the integrated area of the NL transmission curve, between 1 – 1.3 THz [indicated by the dotted vertical lines in Figure 2(a)], compared to the linear excitation case, as shown in figure 2(b). It can be seen that peak transmittance near 1.17 THz increases quadratically with the THz field, which is a signature of an NL phenomenon. As discussed above, the high-frequency spectral content of the driving THz field does not vary much with peak electric fields. In contrast, the low-frequency spectral content increases almost linearly with peak electric fields. Hence, we believe that the enhancement observed in the sample transmission is due to an NL process driven by the lower frequency components in the THz electric field rather than the higher frequency components.

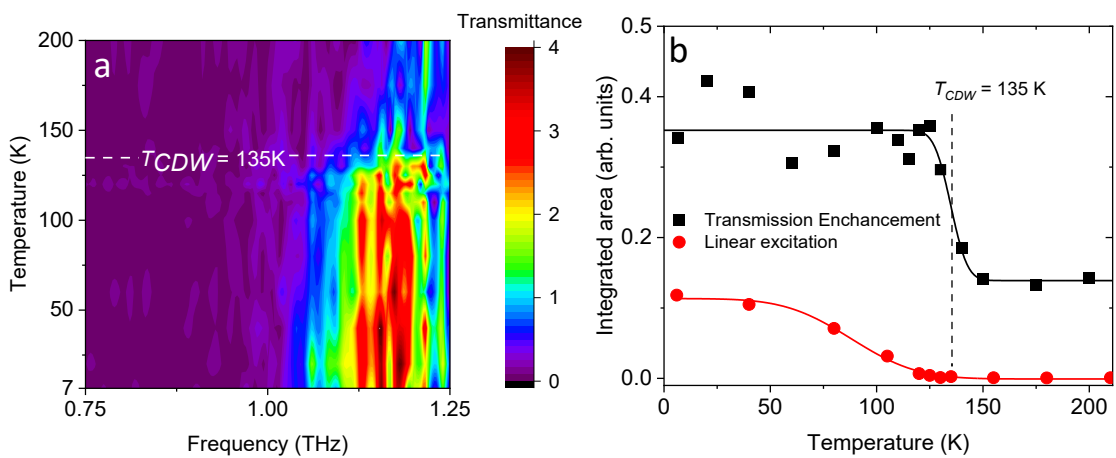

Figure 3. (a) The temperature-dependent transmittance spectra in the nonlinear regime for the SCNO sample. The THz peak electric field was 6 kV/cm. (b) Transmission enhancement is calculated as the difference between NL and linear transmittance, integrated over the region from 1 – 1.3 THz, as a function of temperature. For comparison, the integrated transmission for the same range of frequencies in the linear regime is also shown. The solid lines are guides to the eye.

To further understand the origin of this enhancement, we studied the temperature dependence of the transmittance, which is shown in Figure 3 (a). A sudden increase in the enhancement is observed below $T_{CDW}$, as shown in Figure 3(b), which points to its relation to the CDW formation in the ladders of SCNO. We further studied an undoped SCO sample, and a sharper peak structure near 1.1 THz was observed, as shown in Figure 4. Even though the transmission enhancement near 1.1 THz was not as prominent as the SCNO sample, the peak-like structure indicates that this mode is an intrinsic feature of the SCO system.

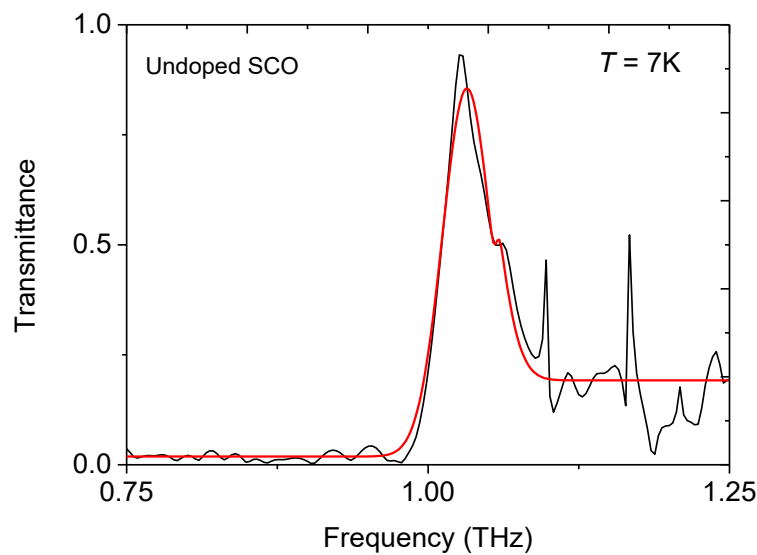

Figure 4. Nonlinear transmittance of the undoped SCO crystal at $T$ = 7 K. The thick red line represents a fit to the mode with a function as described in the main text.

We first look at the various possibilities that could result in such an enhancement in the THz signal. Several different modes have been reported in the SCO system, with frequencies in the sub-THz range. [16,18–20,25,26] However, a mode at ≈ 1.17 THz has not been reported so far in optical spectroscopy measurements, which indicates that the ≈ 1.17 THz mode is an optically silent mode lacking infrared or Raman cross-section. One possibility is the parametric amplification of coherent phasons, similar to what has been reported earlier for the CDW compound DyTe$_3$.[27] Lower frequency phason mode seeded by fluctuations and parametrically amplified by anharmonic coupling to higher frequency amplitudon mode was reported earlier for DyTe$_3$.[27] In the CDW compound K$_{0.3}$MnO$_3$, using time- and angle-resolved photoemission spectroscopy, parametric amplification of the phase fluctuations has been shown to lead to coherent excitation of the Raman-inactive phasons.[28] This amplification was made possible by an anharmonic coupling between the amplitudon and the phason. The amplitude mode is weakly momentum-dependent, while the phase mode energy near the zone center is zero and increases linearly with momentum. However, in the case of a commensurate crystal and pinning due to impurities, the phase mode frequency is non-zero at $q = 0$. In the case of SCO, phasons have been reported earlier at ≈ 0.93 THz. [24] It has also been reported that due to NL coupling, silent goldstone-like phonon modes can be coherently excited by the down-conversion of resonantly driven Higgs modes.[29] The coherent Higgs mode acts as a source of parametric amplification of the phase fluctuation, enhancing the reflectivity of the probe pulse.[30] Further, using MIR radiation, parametric amplification of optical phonons through NL response of the lattice has been reported.[31] Our experimental observation is similar to Cartella et al. [31], where the reflectivity of the probe was reported to be greater than 1, using a higher energy pump, evidencing amplification. However, we discount these parametric downconversion processes discussed above since the spectral component above ≈ 1.17 THz in our driving THz pulse is very low to cause any enhancement by downconversion to 1.17 THz.

Another possibility is a sum-frequency phonon generation process, and it has been shown earlier in other systems that Raman active, infrared forbidden modes at higher frequencies could be excited by intense THz radiation at half its energy.[32] This sum-frequency generation process is different from the parametric amplification, where NL coupling of the dielectric function to the phonon amplitude plays a role, and no phonon-phonon coupling is involved. However, considering our broadband excitation and the observed narrower mode at ≈ 1.17 THz, such a process is unlikely.

As noted earlier, even though several modes in the sub-THz frequency range have been observed in the spin-ladder compound in far-IR studies, a mode at 1.17 THz has not been observed in spectroscopy measurements, indicating that the mode, possibly an optical phonon mode, is silent. Low-lying optical phonon modes with energy ≈ 5 meV (≈ 1.2 THz) have been reported in inelastic neutron scattering studies.[33] It has been shown that a silent mode could be driven indirectly by its coupling to a phonon mode. [14] and such indirect coupling is possible if a higher-order harmonic of the driving mode overlaps with the silent mode. The silent optical phonon mode at 1.17 THz overlaps with the third harmonic of the sliding phonon (IR1) mode at ≈ 0.39 THz, as reported earlier.[20] Hence, when driving with high electric fields, it is possible to obtain an NL coupling of the sliding phonon to the 1.17 THz silent mode. We believe that the enhancement at 1.17 THz is due to its NL coupling of the optical phonon mode with the sliding phonon mode.

In conclusion, we have shown the indirect coupling and upconversion of the sliding phonon mode in the spin-ladder compound to a silent optical phonon mode at higher energy due to nonlinearities induced by the THz electric field. These results open up opportunities to use THz radiation to control high-frequency phonon modes by NL upconversion of lower-frequency phonons.

## Limitations of the study:

We have shown the NL coupling of a silent optical phonon mode to the low energy sliding phonon in the spin-ladder compound, $Sr_{14}Cu_{24}O_{41}$ doped with 1% Ni. Our studies were limited to THz transmission measurements using about 6 kV/cm peak electric fields. Further insights could be obtained by performing THz measurements with higher electric fields and THz pump-probe measurements.

## Acknowledgments

R.N.K. acknowledges the funding support by the Science and Engineering Research Board, Department of Science and Technology, India, through the Core Research Grant No CRG/2019/004865. S.J.S acknowledges support from the Department of Science and Technology, India, through Inspire Fellowship.

## Author contributions

Methodology, S.J.S.; Validation, N.M.; Formal Analysis, S.J.S, N.M.; Investigation, S.J.S, N.M., S.Y.M.; Resources, R.B., S.S., R.N.K.; Data Curation, S.J.S., N.M.; Writing – Original Draft –N.M. S.Y.M., R.N.K.; Writing – Review & Editing, S.J.S, N.M, S.Y.M R.B., S.S, R.N.K.; Visualization, S.J.S., N.M.; Conceptualization, Supervision, Project Administration and Funding Acquisition, R.N.K.

## Declaration of interests

The authors declare that they have no known competing financial interests or personal relationships that could have influenced the work reported in this paper.

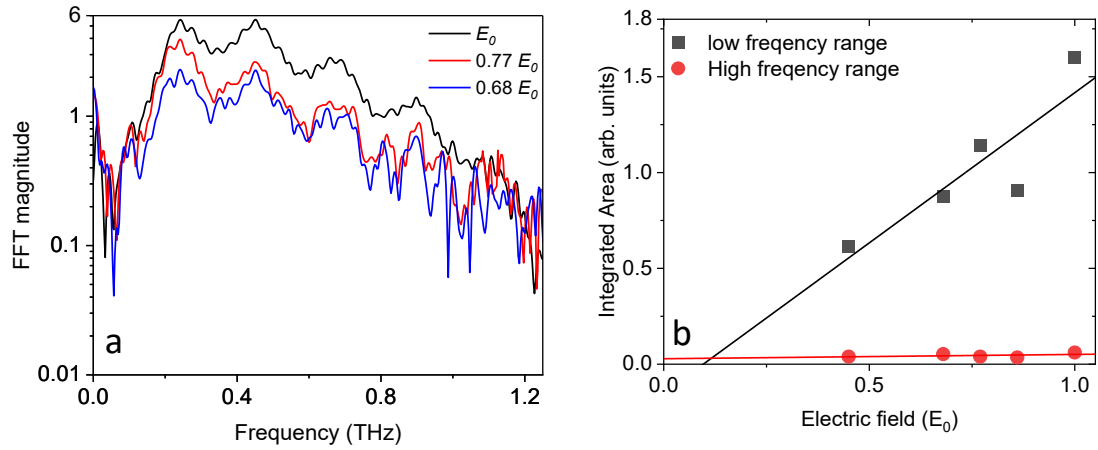

Figure S1. (a) The FFT magnitude of the reference signal ($E_0$) for the NL measurements at different THz peak electric fields and the signal transmitted through the SCNO crystal at 7 K, excited with peak field $E_0$. $E_0$ corresponds to 6 kV/cm. (b) The integrated area of the low-frequency (0.25 – 0.45 THz) region and the high-frequency (1.05 – 1.25 THz) region of the THz spectrum with different peak electric fields.